%Paper: astro-ph/9407071
%From: Brada Rafi <fnbrada@wicc.weizmann.ac.il>
%Date: Sun, 24 Jul 1994 08:41:20 GMT

%plain TeX file
% title: exact solutions and approximations of mond
% fields of disk galaxies.

\magnification=\magstep1
%\input defa
%Article written in TEX.
%\magnification=\magstep1
\font\eighteenrm = cmr10 scaled\magstep3
\font\eighteeni = cmmi10 scaled\magstep3
\font\eighteensy = cmsy10 scaled\magstep3
\font\eighteenit = cmti10 scaled\magstep3
\font\eighteenb = cmbx10 scaled\magstep3

\font\twelverm = cmr12

\font\twelvei = cmmi12
\font\twelveit = cmti12
\font\twelveb = cmbx12
\font\twelvesy = cmsy10 scaled\magstep1
\font\twelves = cmsl12

\font\tenrm = cmr10
\font\tenit = cmti10
\font\teni = cmmi10                  %WICC
\font\tensy = cmsy10                %WICC
\font\tens = cmsl10                  %WICC
\font\tenb = cmbx10                  %WICC
                %WICC

\font\teniu = cmu10
\font\ninerm = cmr9
\font\ninesy = cmsy9
\font\nineb = cmbx9

\font\eightrm = cmr8
\font\eighti = cmmi8
\font\eightsy = cmsy8

\font\sixrm = cmr6
\font\sixi  = cmmi6
\font\sixsy = cmsy6

\font\fivesy = cmsy5

\def\tenpoint{\def\rm{\fam0\tenrm}%                               %WICC
\textfont0=\tenrm \scriptfont0=\eightrm \scriptscriptfont0=\sixrm%WICC
\textfont1=\teni \scriptfont1=\eighti \scriptscriptfont1=\sixi%WICC
\textfont2=\tensy \scriptfont2=\eightsy \scriptscriptfont2=\sixsy%WICC
\textfont3=\tenex \scriptfont3=\tenex \scriptscriptfont3=\tenex
\def\sy{\fam4\tensy}%
\textfont4=\tensy%
\def\sl{\fam5\tens}%
\textfont5=\tens%
\def\bf{\fam6\tenb}%
\textfont6=\tenb%
\def\it{\fam7\tenit}%
\textfont7=\tenit%
\def\prfnt{\fam8\ninesy}%
\textfont8=\ninesy%
\def\hbfnt{\fam9\fivesy}%
\textfont9=\fivesy%
\textfont11=\sixrm \scriptfont11=\sixrm \scriptscriptfont11=\sixrm%
\baselineskip 12pt                                                %WICC
\lineskip 1pt
\parskip 5pt plus 1pt
\abovedisplayskip 12pt plus 3pt minus 9pt
\belowdisplayshortskip 7pt plus 3pt minus 4pt \tenrm}
\def\prfnt{\ninesy }
\def\hbfnt{\fivesy }

\def\bigfnt{\twelverm}

\def\eighteenpoint{\def\rm{\fam0\eighteenrm}%
\textfont0=\eighteenrm \scriptfont0=\twelverm \scriptscriptfont0=\eightrm
\textfont1=\eighteeni \scriptfont1=\twelvei \scriptscriptfont1=\eighti
\textfont2=\eighteensy \scriptfont2=\twelvesy \scriptscriptfont2=\eightsy
\textfont3=\tenex \scriptfont3=\tenex \scriptscriptfont3=\tenex
\def\sy{\fam4\eighteensy}%
\textfont4=\eighteensy%
\def\bf{\fam6\eighteenb}%
\textfont6=\eighteenb%
\def\it{\fam7\eighteenit}%
\textfont7=\eighteenit%
\baselineskip 21pt
\lineskip 1pt
\parskip 5pt plus 1pt
\abovedisplayskip 15pt plus 5pt minus 10pt
\belowdisplayskip 15pt plus 5pt minus 10pt
\abovedisplayshortskip 13pt plus 8pt
\belowdisplayshortskip 10pt plus 5pt minus 5pt
\eighteenrm}
\def\twelvepoint{\def\rm{\fam0\twelverm}%
\textfont0=\twelverm \scriptfont0=\tenrm \scriptscriptfont0=\eightrm
\textfont1=\twelvei \scriptfont1=\teni \scriptscriptfont1=\eighti
\textfont2=\twelvesy \scriptfont2=\tensy \scriptscriptfont2=\eightsy
\textfont3=\tenex \scriptfont3=\tenex \scriptscriptfont3=\tenex
\def\sy{\fam4\twelvesy}%
\textfont4=\twelvesy%
\def\sl{\fam5\twelves}%
\textfont5=\twelves%
\def\bf{\fam6\twelveb}%
\textfont6=\twelveb%
\def\it{\fam7\twelveit}%
\textfont7=\twelveit%
\def\prfnt{\fam8\ninesy}%
\textfont8=\ninesy%
\def\hbfnt{\fam9\fivesy}%
\textfont9=\fivesy%
\def\paperfont{\fam10\twelverm}%
\def\dotfont{\fam11\sixrm}%
\textfont11=\sixrm \scriptfont11=\sixrm \scriptscriptfont11=\sixrm%
\baselineskip 15pt
\lineskip 1pt
\parskip 5pt plus 1pt
\abovedisplayskip 15pt plus 5pt minus 10pt
\belowdisplayskip 15pt plus 5pt minus 10pt
\abovedisplayshortskip 13pt plus 8pt
\belowdisplayshortskip 10pt plus 5pt minus 5pt \twelverm}

\tenpoint
\hsize 5.9truein
\vsize 9.5truein
\global \topskip .7truein
\newcount\chapnum
\chapnum = 0
\newcount\sectnum
\sectnum = 0
\newcount\subsecnum            %17.1.84 mc
\subsecnum = 0        %17.1.84 mc
\newcount\eqnum
\newcount\chapnum
\chapnum = 0
\newcount\sectnum
\sectnum = 0
\newcount\subsecnum            %17.1.84 mc
\subsecnum = 0        %17.1.84 mc
\newcount\eqnum
\eqnum = 0
\newcount\refnum
\refnum = 0
\newcount\tabnum
\tabnum = 0
\newcount\fignum
\fignum = 0
\newcount\footnum
\footnum = 1
\newcount\pointnum
\pointnum = 0
\newcount\subpointnum
\subpointnum = 96
\newcount\subsubpointnum
\subsubpointnum = -1
\newcount\letnum
\letnum = 0
\newbox\referens
\newbox\figures
\newbox\tables
\newbox\tempa
\newbox\tempb
\newbox\tempc
\newbox\tempd
\newbox\tempe
\hbadness=10000
\newbox\refsize
\setbox\refsize\hbox to\hsize{ }
\hbadness=1000
\newskip\refbetweenskip
\refbetweenskip = 5pt
\def\ctrline#1{\line{\hss#1\hss}}
\def\rjustline#1{\line{\hss#1}}
\def\ljustline#1{\line{#1\hss}}

\def\ctr#1{\hfill{#1}\hfill}

\def\spose#1{\hbox to 0pt{#1\hss}}
\def\chskipt{\vskip .125in plus 0pt minus 0pt }              %WICC
\def\chskipl{\vskip .7in plus 18pt minus 10pt}               %WICC
\def\secskipt{\penalty-500\vskip 24pt plus 2pt minus 2pt}    %WICC
\def\secskipl{\vskip 3.5pt plus 1pt }
\def\subsecskip{\penalty-500\vskip 6pt plus 2pt minus 2pt }
\def\unchskip{\vskip -.7in }                                 %WICC
\def\conskip{\vskip 14pt }
\newif\ifoddeven
\gdef\oneside{\oddevenfalse}

\oneside
\newif\ifnonumpageone

\gdef\nonumberfirst{\nonumpageonetrue}
\nonumberfirst
\output{\ifoddeven\leftright\else\samemarg\fi
	\ifnonumpageone\checkpage\else\empty\fi
	\plainoutput}
\def\leftright{\ifodd\count0{\global\hoffset=\oddmargin}
		       \else{\global\hoffset=\evenmargin}\fi}
\def\samemarg{\global\hoffset=\oddmargin}
\gdef\oddmargin{.25truein}
\gdef\evenmargin{0truein}
\def\checkpage{\ifnum\count0=1\nopagenumbers\else\empty\fi}
\footline={{\pagefont\hss--\hquad\folio\hquad--\hss}}
\def\pagefont{\teniu}
\setbox\referens\vbox{\ctrline{\bf References }\chskipt }
\setbox\figures\vbox{\ctrline{\bf Figure Captions }\chskipt }
\setbox\tables\vbox{\ctrline{\bf Table Captions }\chskipt }

\def\title#1{\ctrline {\titfnt #1} }
\def\titfnt{\eighteenpoint}
\def\author#1{\ctrline{\autfnt #1}\par}
\def\autfnt{\bigfnt}

\def\abstract{
\ctrline{\bf ABSTRACT}\chskipt}

\def\reset{\global\sectnum = 0 \global\eqnum = 0
     \global\subsecnum = 0}
\def\chap#1{\global\advance\chapnum by 1 \reset %\chskipt    WICC
\endpage
\chskipt
\ifodd\count0{
\rightline{{\chapnumfont Chapter \the\chapnum}}
\medskip
\rightline{{\chapfont #1}}}
\else{
\leftline{{\chapnumfont Chapter \the\chapnum}}
\medskip
\ljustline{{\chapfont #1}}}\fi
\penalty 100000 \chskipl  \penalty 100000
{\let\number=0\edef\next{
\write2{\bigskip\noindent
  \tofcfont Chapter \the\chapnum.{ }#1
  \leadtofc\number\count0\smallskip}}
\next}}
\def\titcon#1{\unchskip
\ifodd\count0{
\rightline{{\chapfont #1}}}
\else{
\leftline{{\chapfont #1}}}\fi
\penalty 100000 \chskipl \penalty 100000 }
\def\chapnumfont{\tenit}
\def\chapfont{\eighteenpoint}
\def\chapter#1{\chap{#1}}
     %for Facil Tex users
\def\sect#1{\global\advance\sectnum by 1 \global\subsecnum = 0
\secskipt
\ifnum\chapnum=0{{\sectfont\par\noindent
\secsign\the\sectnum{ }{ }#1}\par
	 {\let\number=0\edef\next{
	 \write2{\vskip 0pt\noindent\hangindent 30pt
	 \tofcfont\hbox to 30pt{\hfill\the\sectnum\quad}\unskip#1
	 \leadtofc\number\count0}}
	 \next}}
\else{{\sectfont\par\noindent
\secsign\the\chapnum .\the\sectnum{ }{ }#1}\par
      {\let\number=0\edef\next{
       \write2{\vskip 0pt\noindent\hangindent 30pt
       \tofcfont\hbox to 30pt{\hfill\the\chapnum
	      .\the\sectnum\quad}\unskip#1
	 \leadtofc\number\count0}}
       \next}}\fi
       \nobreak\medskip\nobreak}
\def\sectfont{\twelvepoint}
       %for Facil Tex users
\def\secsign{\S}

\def\subsect#1{\global\advance\subsecnum by 1 \secskipt
\noindent
\ifnum\chapnum=0{
     {\subsecfont\the\sectnum .\the\subsecnum{ }{ }#1}
      \nobreak\par
     {\let\number=0\edef\next{
     \write2{\vskip 0pt\noindent\hangindent 58pt\tofcfont
     \hbox to 60pt{\hfill\the\sectnum
	   .\the\subsecnum\quad}\unskip#1
     \leadtofc\number\count0}}\next}}
\else{{
     \subsecfont\the\chapnum .\the\sectnum
	.\the\subsecnum{ }{ }#1}\nobreak\par
     {\let\number=0\edef\next{
     \write2{\vskip 0pt\noindent\hangindent 58pt\tofcfont
     \hbox to 60pt{\hfill\the\chapnum.\the\sectnum
	   .\the\subsecnum\quad}\unskip#1
     \leadtofc\number\count0}}\next}}\fi
     \nobreak\medskip\nobreak}

\def\subsecfont{\twelvepoint}        %17.1.84 mc
\immediate\openout 2 tofc
\def\tableofcontents#1{\endpage
\count0=#1
\chskipt
\ifodd0{
\rjustline{{\chapfont Contents}}}
\else{
\ljustline{{\chapfont Contents}}}\fi
\chskipl
\rjustline{{\tofcfont Page}}
\bigskip
\immediate\closeout 2
\input tofc
\endpage}

\def\tofcfont{\ninerm}
\def\leadtofc{\leaders\hbox to 8pt{\hfill.\hfill}\hfill}
\immediate\openout 4 refc
\def\refbegin#1#2{\unskip\global\advance\refnum by1
\xdef\rnum{\the\refnum}
\xdef#1{\the\refnum}
\xdef\rtemp{$^{\rnum}$}
\unskip
\immediate\write4{\vskip 5pt\par\noindent\tofcfont
  \hangindent .11\wd\refsize \hbox to .11\wd\refsize{\hfill
  \the\refnum . \quad } \unskip}\unskip
  \immediate\write4{#2}\unskip}

\def\ref#1{\refbegin{\?}{#1}}

\def\refsbegin#1#2{\unskip\global\advance\refnum by 1
\xdef\refb{\the\refnum}
\xdef#1{\the\refnum}
\xdef\rrnum{\the\refnum}
\unskip
\immediate\write4{\vskip 5pt\par\noindent\tofcfont
  \hangindent .11\wd\refsize \hbox to .11\wd\refsize{\hfill
  \the\refnum . \quad } \unskip}\unskip
  \immediate\write4{#2}\unskip}
\def\REFSCON#1#2{\unskip \global\advance\refnum by 1
\xdef#1{\the\refnum}
\xdef\rrrnum{\the\refnum}
\unskip
\immediate\write4{\vskip 5pt\par\noindent\tofcfont
  \hangindent .11\wd\refsize \hbox to .11\wd\refsize{\hfill
  \the\refnum . \quad } \unskip}\unskip
  \immediate\write4{#2}\unskip}

\def\refsend{\nobreak$^{\refb-\the\refnum}$\unskip}

\def\foot#1{\footnote{$^{\the\footnum}$}{#1}
  \global\advance\footnum by 1}

\def\figure#1#2{\global\advance\fignum by 1
\xdef#1{\the\fignum }
\ctrline{\Figure . #2}\par\conskip
{\let\number=0\edef\next{
\write3{\par\noindent\tofcfont
  \hangindent .11\wd\refsize \hbox to .11\wd\refsize{\hfill
  \the\fignum . \quad } \unskip}
\write3{#2\leadtofc\number\count0\par}}
\next}}
\def\figurs#1#2#3{\global\advance\fignum by 1
\xdef#1{\the\fignum }
\ctrline{\Figure . \it #2}
\ctrline{\it #3}\par\conskip
{\let\number=0\edef\next{
\write3{\par\noindent\tofcfont
  \hangindent .11\wd\refsize \hbox to .11\wd\refsize{\hfill
  \the\fignum . \quad}\unskip}
\write3{#2 #3\leadtofc\number\count0\par}}
\next}}

\def\figcon{\ctrline{{\it Figure  \the\fignum} -- cont'd}\par\conskip}
\def\Figure{{\it Figure  \the\fignum}}
\immediate\openout 3 figc

\def\table#1#2{\global\advance\tabnum by 1
\xdef#1{\the\tabnum }
\ctrline{\Table . #2}\par\conskip
{\let\number=0\edef\next{
\write5{\par\noindent\tofcfont
  \hangindent .11\wd\refsize \hbox to .11\wd\refsize{\hfill
  \the\tabnum . \quad } \unskip}
\write5{#2\leadtofc\number\count0\par}}
\next}}
\def\tabls#1#2#3{\global\advance\tabnum by 1
\xdef#1{\the\tabnum }
\ctrline{\Table . #2}
\ctrline{\it #3}\par\conskip
{\let\number=0\edef\next{
\write5{\par\noindent\tofcfont
  \hangindent .11\wd\refsize \hbox to .11\wd\refsize{\hfill
  \the\tabnum . \quad}\unskip}
\write5{#2 #3\leadtofc\number\count0\par}}
\next}}

\def\Table{\it Table  \the\tabnum}
\immediate\openout 5 tabc

\def\eqname#1{\global\advance\eqnum by 1
\ifnum\chapnum=0{
   \xdef#1{ (\the\eqnum ) }(\the\eqnum )  }
\else{
   \xdef#1{ (\the\chapnum .\the\eqnum ) }
	    (\the\chapnum .\the\eqnum ) }\fi}
\def\enum{\global\advance\eqnum by 1
  \ifnum\chapnum=0{ (\the\eqnum )  }
  \else{(\the\chapnum .\the\eqnum ) }\fi}

\def\eqn#1{\eqno \eqname{#1} }
\def\eqnameap#1{\global\advance\eqnum by 1
   \xdef#1{ (\copy\appbox .\the\eqnum ) }
	    (\copy\appbox .\the\eqnum ) }
\def\enumap{\global\advance\eqnum by 1
  (\copy\appbox .\the\eqnum ) }

\def\item#1{\par\noindent\hangindent .08\wd\refsize
\hbox to .08\wd\refsize{\hfill #1\quad}\unskip}
\def\sitem#1{\par \noindent\hangindent .13\wd\refsize
\hbox to .13\wd\refsize{\hfill #1\quad}\unskip}
\def\ssitem#1{\par\noindent\hangindent .195\wd\refsize
\hbox to .195\wd\refsize{\hfill #1\quad}\unskip}

\def\point{\par \global\advance\pointnum by 1
\noindent\hangindent .08\wd\refsize \hbox to .08\wd\refsize{\hfill
\the\pointnum .\quad}\unskip}

\def\spoint{\par \global\advance\subpointnum by 1
\noindent\hangindent .13\wd\refsize
\hbox to .13\wd\refsize{\hfill
(\char\the\subpointnum )\quad}\unskip}

\def\sspoint{\par \global\advance\subsubpointnum by 1
\noindent\hangindent .195\wd\refsize
\hbox to .195\wd\refsize {\hfill\hbox to 20pt{
(\romannumeral\subsubpointnum\hfill)}\quad}\unskip}

\def\bye{\endpage\end}              %for Facil Tex users

\def\bspace#1{\hbox to -#1{}}
\newbox\appbox
\def\appendix#1{\endpage\reset
\setbox\appbox\hbox{#1}
\chskipt \ctrline {\bf APPENDIX #1 }\penalty 10000
\chskipl \penalty 10000
\write2{\bigskip\noindent
 {\tofcfont Appendix #1\leadtofc\number\count0\par\smallskip}}}

\def\mat#1#2{\if 2#1 {\left( \  \vcenter{\halign{$\ctr{## }$ \quad
& $\ctr{## }$\cr #2}} \  \right) } \else{ }\fi
\if 3#1 {\left( \  \vcenter{\halign{
$\ctr{## }$ \quad & $\ctr{## }$ \quad
& $\ctr{## }$\cr #2}} \  \right) } \else{ }\fi
\if 4#1 {\left( \  \vcenter{\halign{$\ctr{## }$ \quad &
$\ctr{## }$ \quad & $\ctr{## }$ \quad
& $\ctr{## }$\cr #2}} \  \right) } \else{ }\fi
\if 5#1 {\left( \  \vcenter{\halign{$\ctr{## }$ \quad
& $\ctr{## }$ \quad & $\ctr{## }$ \quad
& $\ctr{## }$ \quad & $\ctr{## }$\cr #2}} \  \right)} \else{ }\fi
\if 6#1 {\left( \  \vcenter{\halign{$\ctr{## }$ \quad
& $\ctr{## }$ \quad & $\ctr{## }$ \quad & $\ctr{## }$ \quad
& $\ctr{## }$ \quad & $\ctr{## }$\cr #2}} \  \right)} \else{ }\fi }

\def\endpage{\par \vfill \eject}
\def\physrev{\baselineskip 24pt
\lineskip 1pt
\parskip 1pt plus 1pt
\abovedisplayskip 15pt plus 7pt minus 13.33pt
\belowdisplayskip 15pt plus 7pt minus 13.33pt
\abovedisplayshortskip 14pt plus 11pt
\belowdisplayshortskip 9pt plus 7pt minus 7pt
\def\chskipt{\vskip 24pt }
\def\chskipl{\vskip 6.5pt }
\def\secskipt{\vskip 7pt plus 3pt minus 1.33pt }
\def\secskipl{\vskip 3.5pt plus 2pt }
\def\subsecskip{\vskip 7pt plus 2pt minus 2pt }
\def\unchskip{\vskip -6.5pt }
\def\conskip{\vskip 24pt }
\refbetweenskip = \the\baselineskip
\multskip\refbetweenskip by 5
\divskip\refbetweenskip by 10
\twelverm }
\def\bk{\hfil\break}         %WICC
               %for Faci Tex users

\def\To{\par\noindent\hangindent .18\wd\refsize
\hbox to .18\wd\refsize {To: \hfill \qquad }}
\def\from{\par\noindent\hangindent .18\wd\refsize
\hbox to .18\wd\refsize {From: \hfill \qquad }}
\def\topic{\par\noindent\hangindent .18\wd\refsize
\hbox to .18\wd\refsize{Topic: \hfill \qquad }}

\def\startpage#1 {\count0 = #1}
\def\startchapter#1{\chapnum = #1 \advance\chapnum by -1}

\def\startfig#1{\fignum = #1 \advance\fignum by -1}
\def\starttab#1{\tabnum = #1 \advance\tabnum by -1}

\newbox\xa\newbox\xb
\def\boxit#1{ \setbox\xa\vbox {\vskip \boxitsep
\hbox{\hskip \boxitsep #1\hskip \boxitsep }\vskip \boxitsep }
\setbox\xb\hbox{\vrule \copy\xa \vrule}
\vbox{\hrule width 1\wd\xb \copy\xb \hrule width 1\wd\xb }}
\def\fboxit#1#2{ \setbox\xa\vbox {\vskip \boxitsep
\hbox{\hskip \boxitsep #2\hskip \boxitsep }\vskip \boxitsep }
\setbox\xb\hbox{\vrule width #1pt \copy\xa \vrule width #1pt}
\vbox{\hrule height #1pt width 1\wd\xb
\copy\xb \hrule height #1pt width 1\wd\xb }}
\def\reboxit#1#2#3{ \setbox\xa\vbox{\vskip \boxitsep
\hbox{\hskip \boxitsep #3\hskip \boxitsep }\vskip \boxitsep }
\setbox\xb\hbox{\vrule width #1pt\bspace{#2}
\copy\xa \vrule width #1pt}
\vbox{\hrule height #1pt width 1\wd\xb
\copy\xb \hrule height #1pt width 1\wd\xb}}
\def\boxitsep{4pt}

\newdimen\offdimen
\def\offset#1#2{\offdimen #1
   \noindent \hangindent \offdimen
   \hbox to \offdimen{#2\hfil}\ignorespaces}
\newdimen\defnamespace   % Space for name in aligned definition
\defnamespace=2in        % Default value
\def\definition#1#2{     % Aligned definition
    \def\itema{\par\hang\textindent}
    {\advance\parindent by \defnamespace
     \advance\defnamespace by -.5em
     \itema{\hbox to \defnamespace{#1\hfil}}#2\par}}
     %^   YF
\def\TEX{\hbox{T\hskip-2pt\lower1.94pt\hbox{E}\hskip-2pt X}}
\def\wyz{\hbox{WI\hskip-1pt\lower.9pt\hbox{Z\hskip-1.85pt
\raise1.7pt\hbox{Z}}LE}}
    %cents sign
\def\\{$\backslash $}%\ when not in MM
     % \lc & \rc make braces when not in MM

\def\underwiggle#1{\mathop{\vtop{\ialign{##\crcr
    $\hfil\displaystyle{#1}\hfil$\crcr\noalign{\kern2pt\nointerlineskip}
    $\scriptscriptstyle\sim$\crcr\noalign{\kern2pt}}}}\limits}
\def\({[}
\def\){]}

\def\hquad{\hskip.5em{}}
\mathchardef\app"3218

\def\linebreak{\break}
\mathchardef\oprod="220A
\mathchardef\inter="225C
\mathchardef\union="225B

\mathchardef\relv="326A
\mathchardef\leftv="326A
\mathchardef\rightv="326A
\mathchardef\relvv"326B
\mathchardef\leftvv"326B
\mathchardef\rightvv"326B
\mathchardef\Zscr"25A
\mathchardef\Yscr"259
\mathchardef\Xscr"258
\mathchardef\Wscr"257
\mathchardef\Vscr"256
\mathchardef\Uscr"255
\mathchardef\Tscr"254
\mathchardef\Sscr"253
\mathchardef\Rscr"252
\mathchardef\Qscr"251
\mathchardef\Pscr"250
\mathchardef\Oscr"24F
\mathchardef\Nscr"24E
\mathchardef\Mscr"24D
\mathchardef\Lscr"24C
\mathchardef\Kscr"24B
\mathchardef\Jscr"24A
\mathchardef\Iscr"249
\mathchardef\Hscr"248
\mathchardef\Gscr"247
\mathchardef\Fscr"246
\mathchardef\Escr"245
\mathchardef\Dscr"244
\mathchardef\Cscr"243
\mathchardef\Bscr"242
\mathchardef\Ascr"241
\mathchardef\lscr"160

\immediate\openout 2 tofc
\immediate\openout 3 figc
\immediate\openout 4 refc
\immediate\openout 5 tabc
\tolerance 4000
%---Puts reference numbers on line
\def\refbegin#1#2{\unskip\global\advance\refnum by1
\xdef\rnum{\the\refnum}
\xdef#1{\the\refnum}
F\xdef\rtemp{[\rnum]}
\unskip
\immediate\write4{\vskip 5pt\par\noindent\tofcfont
  \hangindent .11\wd\refsize \hbox to .11\wd\refsize{\hfill
  \the\refnum . \quad } \unskip}\unskip
  \immediate\write4{#2}\unskip}
\def\refsend{\nobreak[\refb-\the\refnum]\unskip}
%---------------------
\tolerance 4000
%---Puts reference numbers on line
\def\refbegin#1#2{\unskip\global\advance\refnum by1
\xdef\rnum{\the\refnum}
\xdef#1{\the\refnum}
\xdef\rtemp{[\rnum]}
\unskip
\immediate\write4{\vskip 5pt\par\noindent\tofcfont
  \hangindent .11\wd\refsize \hbox to .11\wd\refsize{\hfill
  \the\refnum . \quad } \unskip}\unskip
  \immediate\write4{#2}\unskip}
\def\refsend{\nobreak[\refb-\the\refnum]\unskip}
% end of defa

\def\rar{\rightarrow}
\def\_#1{_{\scriptscriptstyle #1}}
\def\&#1{^{\scriptscriptstyle #1}}
\def\pd#1#2{{\partial#1\over\partial#2}}

\def\nineb{\bf}
\def\ninerm{}
\def\({[}
\def\){]}

\def\dtv{d\&3 v}
\def\div{\vec\nabla\cdot}
\def\grad{\vec\nabla}
\def\curl{\vec\nabla\times}
\def\ao{a_{o}}

\def\fn{\varphi\_{N}}
\def\fk{\varphi\_{K}}
\def\pk{\psi\_{K}}
\def\abs#1{\vert #1\vert}
\def\ags{\abs{\grad\psi}}
\def\agf{\abs{\grad\fn}}
\def\Skr{\Sigma\_{K}(r)}
\def\l{\lambda}
\def\d{\delta}
\def\a0{a\_{o}}
\vsize 23.6truecm
\hsize 5.6truein
\parskip 0pt
\def\oddmargin{.3in}
\def\evenmargin{.1in}

\def\rhsq{(r\&{2}+h\&{2})}

\def\bk{\par\noindent}
\baselineskip 20pt
\vskip 0.5truein
\centerline{{\bf EXACT SOLUTIONS AND APPROXIMATIONS OF}}
\centerline{{\bf MOND FIELDS OF DISK GALAXIES}}
\vskip 0.5truein
\centerline{ Rafael Brada {\it and}  Mordehai Milgrom}
\bk
\centerline{Department of Condensed-Matter
 Physics, Weizmann Institute of Science}
\centerline{ 76100 Rehovot, Israel}
\vskip 0.5truein
\vskip 30pt
\baselineskip 13pt
\ninerm
{\bf \noindent Abstract}
We Consider models of thin disks (with and without bulges) in the
Bekenstein-Milgrom formulation of MOND as a modification of Newtonian
gravity. Analytic solutions are found for the full gravitational fields
of Kuzmin disks, and of disk-plus-bulge generalizations of them.
  For all these models a simple
 relation between the MOND potential field, $\psi$,
 and the Newtonian potential, $\fn$, holds
everywhere {\it outside the disk}:
$\mu(\ags/\a0)\grad\psi=\grad\fn$. We give exact expressions for
the rotation curves for these models.
We also find that this algebraic relation is a very good approximation
for exponential disks. The algebraic relation outside
the disk is then extended into the disk to
 derive an improved approximation for the MOND rotation curve
of disk galaxies that requires only knowledge of the Newtonian
curve and the surface density.

\vskip 20pt
{\nineb I. Introduction}
\vskip 5pt
\par
There are two extreme interpretations of the modified
Newtonian dynamics (MOND). One of these views
MOND as a modification of inertia (Milgrom 1983a, 1994a):
 Gravitational fields of massive bodies remain Newtonian,
but the equation of motion of a particle in the field is superseded
by a MOND equation of motion.
In this paper, however, we concentrate on the Bekenstein-Milgrom (BM)
 formulation of MOND (Bekenstein and Milgrom 1984, hereafter BM), which
is an embodiment of MOND as a modification of gravity, leaving the
Newtonian law of motion intact.
The standard Poisson equation for the Newtonian gravitational
potential, $\fn$, ($\div\grad\fn=4\pi G\rho$) induced by a mass density
$\rho(\vec R)$ is replaced by
$$\div[\mu(\ags/\a0)\grad\psi]=4\pi G\rho, \eqn{\Ii} $$
with $\a0$ the acceleration constant of MOND.
This non-linear equation is hardly amenable to analytic solution
beyond the simple cases of configurations with one-dimensional
symmetry.
\par
It would be very useful, for example,
 to have exact, or even approximate, analytic
 solutions for the gravitational field of model disk galaxies
on which various ideas can be tested. Some problems whose study may
benefit from the availability of such solutions are, for example,
that of polar rings, and that of the motion and fate (disruption,
capture etc.) of dwarf companions moving in the field of a mother
galaxy.
\par
 Even more central is the problem of calculating the MOND
 rotation curves of disk galaxies.
 In formulations of MOND based on modification of inertia, the
velocity on a circular orbit of radius $r$ in the plane of disk galaxies
is given exactly by
$$\mu(a/\a0)a=a\_{N},   \eqn{\Iii} $$
 where $a=v\&{2}/r$, and $a\_{N}$
is the Newtonian acceleration at $r$ (Milgrom 1994a).
This has been the standard expression for calculating MOND rotation
 curves (e.g. Kent 1987, Milgrom 1988, Begeman, Broeils, and sanders
1991). It is not exact in the Bekenstein-Milgrom formulation, and
had had the status of only an approximation before the work of Milgrom
1994a.
\par
Here we describe a class of disk-galaxy models for which exact
solutions of the MOND field equation are presented; this is done in
 \S III.
We also find (see \S IV)
 that an approximate analytic solution applies for a wider class
of models, and we suggest a way to predict the adequacy of such an
approximation, by studying only the Newtonian solution for the
mass distribution (\S II).
In \S V we describe an approximation for the rotation curve
in the BM formulation--based, like relation\Iii ,
only on knowledge of the Newtonian acceleration, but which is, generally,
a better approximation.
In \S VI we mention further possible developments.

\vskip 20pt
{\nineb II. An algebraic relation between the Newtonian and MOND fields}
\vskip 5pt
\par
Subtracting the Poisson
 equation from the MOND equation\Ii we get
$$\div[\mu(\ags/\a0)\grad\psi-\grad\fn]=0, \eqn{\IIexi} $$
by which the expression in parentheses is some curl field.
  For configurations with one-dimensional symmetry
(spherical, cylindrical, or plane) the curl field must vanish, and thus
 the MOND field is
related to the Newtonian field by
the algebraic relation
$$\mu(\ags/\a0)\grad\psi=\grad\fn.  \eqn{\IIi} $$
This affords a simple solution of the MOND problem by solving first
the Poisson equation for $\fn$,
 then inverting eq.\IIi to get the MOND field.
Relation\IIi does not follow from the MOND equation\Ii, but the inverse
is correct as the latter is just the divergence of the former.
\par
We begin by asking whether such a relation may hold for more general
mass distributions, at least approximately.
 Because the function $\mu$
that appears in MOND  is such that $I(x)\equiv x\mu(x)$
is monotonic, and varies between 0 and $\infty$ as $x$ does so,
 $I(x)$ is invertible on the positive real axis.
Equation\IIi is thus equivalent to
$$\grad\psi=\nu(\agf/\ao)\grad\fn,  \eqn{\IIii} $$
where $\nu(y)\equiv I\&{-1}(y)/y$.
A potential $\psi$ that satisfies this equation exists if and only if
the curl of the right-hand side vanishes, or, in other terms
$$\grad\agf\times\grad\fn=0 \eqn{\IIiia} $$
 (as $\nu '\not=0$).
This, in turn, is tantamount to $\agf$ being some function of $\fn$
$$\agf=f(\fn).   \eqn{\IIiii} $$
We find then a necessary and sufficient
 condition for eq.\IIi to hold for some $\psi$
and some $\mu$ (with $\mu'\not=0$); the condition is
expressed solely in terms
of the {\it Newtonian} field of the given mass distribution.
By eq.\IIi the equipotentials for $\psi$ and $\fn$ coincide, and $\psi$
is thus a function of $\fn$.
\par
A potential $\psi$ that satisfies eq.\IIi in some domain $D$
 is {\it the} MOND solution to the problem
only if $\psi$ also satisfies the correct boundary conditions.
If the sphere at infinity is part of the boundary of $D$, then
$\psi$ automatically satisfies the correct boundary
condition there. The same is true of the jump condition
across a thin sheet of mass. If $\fn$ satisfies it than a $\psi$ that
obeys eq.\IIi (outside the mass sheet)
 satisfies the correct jump condition as well.
\par
Concentrate now on mass distributions that model disk galaxies: An
axisymmetric distribution, symmetric also about a mid-plane, made of
a thin disk of surface density $\Sigma(r)$, and some
bulge-like component. By the above arguments if $\psi$ satisfies
eq.\IIi everywhere outside the disk it is {\it the} MOND solution of the
problem: the boundary conditions are now satisfied automatically by
a solution of eq.\IIi .
 At infinity $\grad\psi\rar(MG\a0)\&{1/2}\vec R/R\&2$,
and just outside the surface of the disk
$$\mu(\ags/\a0)\partial\_{n}\psi=\pm 2\pi\Sigma(r),  \eqn{\IIviii}$$
where $\partial\_{n}$ is te normal component of the gradient.
\par
To assess the applicability of the algebraic relation for a
given configuration we only have to find the Newtonian
potential, and plot $\agf~vs.~\fn$ for points outside the disk.
If the points fall on a line, i.e. if $\agf$ is a function of $\fn$
(a highly non-generic case) than $\grad\psi$ as given by eq.\IIi
is the exact MOND acceleration field {\it outside the disk}.
If $\agf$ and $\fn$ are correlated with only a little scattering,
equation\IIi gives a good approximation to the MOND field (see \S IV for
examples).

\vskip 20pt
{\nineb III. Exact solutions for Kuzmin disks and generalizations
thereof}
\vskip 5pt
\par
The two-parameter family of Kuzmin disks is described by a
Newtonian gravitational potential
$$\fk=-MG/[r\&{2}+(\abs{z}+h)\&{2}]\&{1/2} \eqn{\IIIii} $$
 (see e.g. Binney and Tremaine 1987),
where we use cylindrical coordinates $r,z$.
The potential above the disk ($z>0$) is that of a point mass $M$
placed on the lower $z$-axis at $-\vec h \equiv(0,0,-h)$;
the potential below the disk is produced by the same mass
oppositely placed at $\vec h$. The surface density,
 $\Skr$, matches the jump in the $z$-gradient
of the potential.
$$\Skr=(2\pi G)\&{-1}\pd{\fk}{z}\biggr\vert\_{z=0\&{+}}=
Mh/2\pi\rhsq\&{3/2}.    \eqn{\IIIiii} $$
Everywhere outside the disk the equipotential surfaces are
concentric spheres centered at $\pm\vec h$. Equations \IIiia\IIiii
are thus satisfied
(in this case $\agf=\fn\&{2}/MG$),
 and, by the arguments of \S II, the exact MOND solution
for Kuzmin disks is given, outside the disk, by the algebraic relation
 eq.\IIi .
Thus, {\it outside the disk},
$$\vec g=-\grad\psi=\a0 I\&{-1}(g\_{N}/\a0)\vec g\_{N}/g\_{N},
 \eqn{\IIIiv} $$
where
$$\vec g\_{N}=-MG(\vec R\pm\vec h)/\abs{\vec R\pm\vec h}\&{3}
 \eqn{\IIIv} $$
is the Newtonian acceleration field above (+), and below ($-$) the disk.
The MOND solution, above the disk,
 is simply that of a point mass located at $-\vec h$.

 For very-low-acceleration Kuzmin disks (with $MG/h\&{2}\ll \a0$) we have
$\mu(x)\approx x$, so $I\&{-1}(x)\approx x\&{1/2}$. Then, the
MOND potential is
$$\pk \approx(MG\a0)\&{1/2}ln[r\&{2}+(\abs{z}+h)\&{2}]\&{1/2},
 \eqn{\IIIvi} $$
which can be obtaind by direct integration of
 eq.\IIIiv .
The MOND rotation curve of a Kuzmin disk is, by eq.\IIIiv
$$v\&{2}(r)=\a0 I\&{-1}[g\_{N}(r,0\&{+})/\a0]r\&{2}
/\rhsq\&{1/2},\eqn{\IIIvii} $$
where $g\_{N}(r,0\&{+})=MG/\rhsq$.
It is clear then that we can write
$$v\&{2}(r)=v\&{2}\_{\infty}\eta(\zeta,u), \eqn{\IIIviia} $$
where $v\_{\infty}\equiv(MG\ao)\&{1/4}$ is the asymptotic rotational
 speed, $\zeta\equiv MG/h\&{2}\ao$ is a measure of how deep in the MOND
regime we are, and $u\equiv r/h$.
If we take, for instance, $\mu(x)=x/(1+x\&{2})\&{1/2}$,
then $I\&{-1}(y)=[y\&{2}/2+(y\&{2}+y\&{4}/4)\&{1/2}]\&{1/2}$,
and
$$v\&{2}(r)=v\&{2}\_{\infty}{u\&{2}\over 1+u\&{2}}\left\{\left[1+
{\zeta\&{2}\over 4(1+u\&{2})\&{2}}\right]\&{1/2}+{\zeta\over 2(1+u\&{2})
}\right\}\&{1/2}. \eqn{\IIIviib} $$
\par
 In the limit of very-low-acceleration disks, $\zeta\rar 0$,
one has, {\it independently of the exact form of $\mu(x)$},
$$v\&{2}(r)=v\&{2}\_{\infty}u\&{2}/(1+u\&{2}),\eqn{\IIIviii}$$
as in this limit $I\&{-1}(y)=y\&{1/2}$.
\par
Milgrom (1994b) has proved a virial-like relation for self-gravitating,
 low-acceleration systems, in the BM formulation.
 For thin disks this relation reads (Milgrom 1994b)
$${2\over 3}M\&{3/2}(G\a0)\&{1/2}=\int\_{0}\&{\infty}2\pi r\Sigma(r)
v\&2(r)~dr,  \eqn{\IIIix} $$
where $v(r)$ is the circular rotation speed;
it can be readily verified to hold for the pair of $\Sigma(r)$
and $v(r)$ given by eqs.\IIIiii and \IIIviii respectively.
\par
Kuzmin disks may be generalized into a family of disk-plus-bulge
models that are exactly solvable in MOND. These may be generated
in several equivalent ways.
\par
  For example, beginning with the Newtonian
 potential of a Kuzmin disk $\fk(\vec R)$
we define a new mass distribution whose Newtonian potential is
$$\varphi=U(\fk).  \eqn{\IIIixa}$$
We choose U such that $U(x)\rar x$ for $x\rar 0$; thus, at spatial
infinity $\varphi$ has the same behavior as $\fk$, and satisfies the
correct boundary behavior for a potential of a mass $M$.
The potential $\varphi$ is produced, outside the disk,
 by a mass distribution
$$\rho(\vec R)=(4\pi G)\&{-1}\nabla\&{2}\varphi=
(4\pi G)\&{-1}U''(\fk)(\grad\fk)\&{2},  \eqn{\IIIx} $$
where we have made use of the fact that $\nabla\&{2}\fk=0$.
  From eq.\IIIx , the equidensity surfaces coincide with the
equipotential surfaces (common to $\varphi$ and $\fk$), because
$(\grad\fk)\&{2}$ is a function of $\fk$.
\par
In addition, a disk is needed at $z=0$, with surface density
$$\Sigma(r)=(2\pi G)\&{-1}\pd{\varphi}{z}\biggr\vert\_{z=0\&{+}}=
U'[\fk(r,0)]\Skr, \eqn{\IIIxi} $$
with
$\fk(r,0)=-MG/\rhsq\&{1/2}$.
  For $\rho$ to be non-negative we must have $U''\ge 0$; thus, $U'$
is an increasing function. Since the maximum value of $\fk$ is 0,
and there $U'=1$, we have $U'\le 1$ everywhere, or $\Sigma(r)\le\Skr$.
The total mass (bulge plus disk) contained within an
 equipotential surface $\fk$ is
$$M(\fk)=U'(\fk)M\_{K}(\fk),   \eqn{\IIIxii} $$
where $M\_{K}$ is the mass within $\fk$
 for the generating Kuzmin disk.
This can be seen by applying the Gauss theorem to the equipotential
surface.
\par
All the potentials $\varphi$ defined by eq.\IIIixa satisfy
eqs.\IIiia\IIiii
because $\fk$ does. Thus the algebraic relation\IIi gives the MOND
 solutions for all these model galaxies in term of the Newtonian field
$\grad\varphi=U'(\fk)\grad\fk$.
\par
A different approach, which generates the same family of solvable models,
starts with some spherical density distribution that is centered
at $-\vec h$: $\rho(\vec R)=
\hat\rho(q),~q\equiv[r\&{2}+(\abs{z}+h)\&{2}]\&{1/2}$.
Take the MOND potential in the $z>0$ region to coincide with that of
$\rho(\vec R)$. In the $z<0$ region the potential is defined
 symmetrically. For spherical systems the MOND field is related to
the Newtonian field by the algebraic relation\IIi . Thus, this is also
the case for the model under construction. The ``bulge'' density
that produces the potential is just the part of the spherical
density distribution $\rho(\vec R)$ that is above the mid-plane;
we can dictate it at will. A disk with surface density $\Sigma(r)$
must supplement the bulge to match the jump in the z-gradient.
 If $M(q)\equiv\int\_{0}\&{q}4\pi \l\&{2}\hat\rho(\l)d\l$
 is the spherical mass
 within distance $q=\rhsq\&{1/2}$ from the centre of $\rho(\vec R)$, then
$$\Sigma(r)={M(q)h\over 2\pi q\&{3}}.  \eqn{\IIIxiii}$$
$\Sigma(r)$ is just the surface density
 of a Kuzmin disk with the same $h$ and a mass
equal to the total spherical mass within the sphere going through
the point at $r$ on the disk.
Comparing with eq.\IIIxi we find the corresponding
 $U'(\fk)=M(q)/M(\infty)$, with $q=-MG/\fk$.
\par
A third approach, which we shall not detail here, is to start with the
MOND potential for the Kuzmin disk, $\psi\_{K}$, and construct new
potentials $\psi=S(\psi\_{K})$.
\par
We reiterate that in all the above models, the bulge equidensity surfaces
coincide with equipotential surfaces of the model. This means that
we can readily construct, for the bulge, distribution functions with
isotropic velocity distributions.
These are of the form $\hat f(E)$, with $E=v\&{2}/2+\psi(\vec r)$,
 for which $\rho(\vec r)=\int \dtv \hat f(E)=F[\psi(\vec r)]$.

\vskip 20pt
{\nineb IV. Some other disk-galaxy models}
\vskip 5pt
\par
How good an approximation is the algebraic relation in general?
There clearly are disk models for which it fails rankly. Consider, for
example a disk whose surface density vanishes at the centre. Then,
$\agf$ vanishes both near the centre, and at infinity, while the
 potential, which vanishes at infinity, is non-zero at the
 centre. Thus $\agf$ is anything but a function of $\fn$, and the
algebraic approximation must break appreciably.
\par
We have found that for the very pertinent case of a disk with
an exponential surface-density law, $\Sigma(r)=\Sigma\_{0}exp(-r/h)$,
 the algebraic approximation
holds very well. A disk for which it holds less well is the so-called
Kalnajs disk (characterized by a constant angular velocity on circular
orbits inside the material disk), whose surface density is
 $\Sigma(r)=\Sigma\_{0}[1-(r/h)\&{2}]\&{1/2}$.
We now discuss these two examples in more detail.
\par
As explained in \S II , to be able to foretell the quality of the
algebraic approximation for a given disk, it is enough to look at
the tightness of the relation $\agf~vs.~\fn$.
In Fig. 1 we show this relation for the above two surface-density
distributions, as obtained from numerical calculations using
a multigrid scheme. The code is capable of solving the nonlinear
MOND equation, and is described in detail in Brada
 1994 (in preparation). For reference we also show in Fig. 1a
 the numerical
 results for the Kuzmin disk, which show that the numerical scattering
about the expected exact relation, $\agf=\fn\&{2}/MG$ (marked by
 crosses),  is quite negligible (the slight departure from the exact
 relation is numerical, and stem from the cutoff in the disk at the end
of the mesh). The correlation for the exponential disk (Fig. 1b)
 is also rather tight (but does not follow the asymptotic relation).
 We thus expect the algebraic approximation to be rather good
for the MOND field, {\it for all values of the mean acceleration}.
We plot in Fig. 2 the relative departure, $\d$,
 from the algebraic relation:
$$\vec \d\equiv{\mu(\ags/\a0)\grad\psi-\grad\fn\over \agf}. \eqn{\IVi}$$
  For a very-low-acceleration Kuzmin disk we see that $\d\approx 0$
 everywhere, as expected. For an
 exponential disk in the same limit  ($\Sigma\_{0}\ll \a0/G$),
we see that $\abs{\vec \d}\ll 1$ everywhere outside the disk, in keeping
with the tight $\agf~vs.~\fn$ relation.
  For the Kalnajs disk, we see in Fig. 1c that the $\agf~vs.~\fn$
 relation has rather more scattering, and indeed the plot of $\vec \d$,
 shown in Fig. 2c (again for $\Sigma\_{0}\ll \a0/G$),
 evinces more substantial departure from the algebraic
approximation. An exponential disk with a hole within one-and-a-half
 scale lengths is even a more extreme case described in Figs. 1d and 2d.

\vskip 20pt
{\nineb V. Rotation curves based on the algebraic approximation}
\vskip 5pt
\par
If the algebraic relation\IIi
 holds outside the disk it cannot be correct in the mid-plane of the
thin disk; so, we cannot use eq.\Iii [$\mu(a/\a0)a=a\_{N}$] to
obtain the rotation curve of the model galaxy. Rather, we have to follow
the following procedure:
We need the radial acceleration, $a\_{r}$, in
the mid-plane of the disk. As the acceleration component parallel to
the disk is continuous across the thin disk, $a\_{r}$ is the same as
$a\_{r}\&{+}$, the radial acceleration just outside the disk. This can
be obtained from the algebraic relation in terms of the total
 Newtonian acceleration just outside the disk $a\_{N}\&{+}$, and its
radial component. The latter can again be equated to its value in the
mid-plane of the disk (as it too is continuous), and so we obtain
$$v\&{2}(r)/r=a\_{r}=a\_{r}\&{+}={a\_{rN}\over\mu(a\&{+}/\a0)}=
{a\_{rN}\over\mu[I\&{-1}(a\_{N}\&{+}/\a0)]}.  \eqn{\Viii} $$
To complete the expression we express $a\_{N}\&{+}$ in terms of
Newtonian radial acceleration in the mid-plane, directly related to the
Newtonian rotation curve
 $a\_{N}\&{+}=[a\_{rN}\&{2}+(2\pi G\Sigma)\&{2}]\&{1/2}$.
The MOND rotation curve is
 thus given by a simple function the corresponding Newtonian quantity.
The correction to eq.\Iii involves the addition of the $2\pi G\Sigma$
term in the argument of $I\&{-1}$ in eq.\Viii .
\par
Relation \Iii was found numerically (Milgrom 1986) to constitute a good
approximation for a large class of bulge-plus-disk galaxy models,
 but, as we said, it is
not exact in the BM formulation, even for configurations
for which it is correct outside the disk.
 For example, for the low-surface-density Kuzmin disk, relation\Iii gives
for the rotation speed
$$v\&{2}(r)=(MG\a0)\&{1/2}r\&{3/2}/\rhsq\&{3/4}, \eqn{\Vii} $$
 to be compared with the
somewhat different exact expression\IIIviii\
[ where the $r$ dependence is $r\&{2}/\rhsq$].
 This latter is obtained from eq.\Viii .
\par
 We suggest that Eq.\Viii is, generically,
a better approximation for the rotation curve of disk galaxies in the
BM formulation than is eq.\Iii  even when the algebraic approximation is
not so good outside the disk  (see some examples below);
it is as easy to apply as the latter.
 [We remind the reader that in the
 formulation of MOND as a modification of inertia (Milgrom 1994a)
 relation\Iii gives the rotation curve exactly.]
\par
We give in Fig. 3 three rotation curves for each of a few galaxy models.
The galaxy models presented are the bare Kuzmin disk,
 a bare exponential disk, and a Kalnajs disk, all in the deep MOND limit.
We give the exact rotation curve calculated numerically,
 the curve calculated from the approximate expression
\Iii, and that calculated from what we propose as an improved
 approximation\Viii. We expect the performance of eq.\Iii to be the worst
for pure disks in the deep MOND limit; adding a spherical component,
and/or going nearer the Newtonian regime can only improve its performance
(but not that of approximation \Viii).
\par
Interestingly, the MOND rotation curve for a Kalnajs disk seems to be
given exactly by $v\propto r$--as in the Newtonian limit--no matter how
deep in the MOND regime we are. We do not yet understand the
origin of this behavior. Once this is taken as fact, the proportionality
factor, i.e., the constant angular velocity, $\Omega$,
 may be calculated, for very-low-acceleration disks
 from the virial relation \IIIix for disks to get
$\Omega\&{2}=5\cdot 3\&{-3/2}(2\pi\Sigma\_{0}G\ao)\&{1/2}/h$.
compared with the Newtonian angular velocity which is
$\Omega\_{N}\&{2}=\pi\&{2}G\Sigma\_{0}/2h$.

\vskip 20pt
{\nineb VI. Discussion}
\vskip 5pt
\par
We have described models of disk galaxies for which exact solutions
of the Bekenstein-Milgrom field equation can be obtained in the form
of a simple algebraic relation between the MOND solution, and the
Newtonian field of the same mass distribution.
This relation holds approximately for a wider class of
 configurations, which include exponential disks.
 We have given a simple criterion to assess the
validity, or near validity of this relation; the use of this criterion
assumes knowledge of the Newtonian field, $\fn$, only: it requires that
$\agf$ be tightly correlated with $\fn$
outside the disk. We have also suggested an improved
 approximation--inspired by the above approximation--for
 calculating rotation curves in the BM formulation.
\par
When accuracy beyond the algebraic approximation is needed it
 may serve a first approximation around
which we can linearize the MOND equation in the small increment.
We may, for instance, proceed as follows:
Suppose that the $\agf~vs. ~\fn$ has some scattering but we can
reasonably
define a mean relation $\agf\approx f(\fn)$. The acceleration field
that is derived from the algebraic relation, which is to serve as our
zeroth order approximation, is not derivable from a potential, in
 general. So, it is more convenient to work with accelerations, not
with potentials. Define then
$$\vec q\equiv\mu(\ags/\ao)\grad\psi,  \eqn{\Vi} $$
which is inverted, as in eq.\IIii, to give
$$\grad\psi=\nu(q/\ao)\vec q,  \eqn{\Vii}$$
where $q=\abs{\vec q}$. The algebraic relation would equate
$\vec q$ to $\grad\fn$, but we now write
$$\vec q=\grad\fn+\vec\eta ,  \eqn{\Viii}$$
with $\vec\eta=\agf\vec\d$ a curl field which is
 assumed to be small compared with $\grad\fn$, and which we shall treat
to first order. By the MOND equation we now have
$$\div\vec\eta=0,  \eqn{\Viv}$$
and from eq\Vii
$$\curl[\nu(q/\ao)\vec q]=0.  \eqn{\Vv} $$
Equations\Viv\Vv are equivalent to the original MOND equation for the
potential $\psi$ (see Milgrom 1986).
We now substitute eq.\Viii in eq.\Vv and take the first order in
  $\vec\eta$
(noting that $\grad\agf\times\grad\fn$, which measures the departure
from the algebraic relation, is also first order) to get
$$\curl\vec\eta+\hat\nu\vec e\times[\grad(\vec e\cdot\vec\eta)
-f'(\fn)\vec\eta]=\hat\nu\grad\agf\times\vec e, \eqn{\Vvi} $$
Here $\vec e(\vec r)\equiv -\grad\fn/\agf$ is a unit vector in the
direction of the local Newtonian acceleration, and $\hat\nu(\vec r)$
is the logarithmic derivative of $\nu$ calculated at $\agf/\ao$
 ($\hat\nu=1$ in the deep MOND limit).
The linear equations\Viv\Vvi determine $\vec\eta$.

\par
Our construction of the solvable disk models began with a known MOND
solution which does not involve a disk, such as a point mass, or, in
general, a spherical mass distribution. We than place that mass
 distribution anywhere relative to the $z=0$ plane; then we take
the MOND potential, above the plane only, to be that of the mass
in question, defining the potential below the plane as the mirror
image of the one above.
The disk is then found which matches the jump of the $z$-gradient of the
potential; hence a family of solvable disk models is born.
Clearly, we may start with any axisymmetric
 mass distribution for which the MOND
solution is known analytically or numerically--not just a spherical
 one--and get a new family of disk models.
Such initial non-disk MOND solutions can be found by starting from a
potential, then calculating the density distribution from eq.\Ii ,
making sure that the resulting $\rho$ is positive everywhere, and is
otherwise reasonable.

\endpage
{\bf References}
\vskip 4pt
\bk
Begeman, K.G., Broeils, A.H., and Sanders, R.H., 1991,
MNRAS, 249, 523
\bk
Bekenstein, J., and Milgrom, M., 1984, ApJ, 286, 7 (BM)
\bk
Binney, J., and Tremaine, S., 1987, {\it Galactic Dynamics}, Princeton
University Press, Princeton.
\bk
Kent, S.M., 1987, AJ, 93, 816
\bk
Milgrom, M., 1983a, ApJ, 270, 365
\bk
-----------, 1986, ApJ, 302, 617
\bk
-----------, 1988, ApJ, 333, 689
\bk
-----------, 1994a, Annals of Physics, 229, 384
\bk
-----------, 1994b, ApJ, 429, in the press.
\endpage
   \bk
{\bf Figure captions} \bk
 Figure 1. Plots of $\agf~vs.~\fn$ for a Kuzmin (a), exponential (b), and
Kalnajs (c) disks, and for an exponential disk cutoff below
 one-and-a-half scale lengths.
 The crosses mark the relation $\agf=\fn\&{2}/MG$.
\bk
 Figure 2. A plot of $\vec\d$--a measure of the departure from the
algebraic relation--for the four disks as in Fig. 1.
\bk
 Figure 3. The rotation curves for the first three disk models of
 Fig. 1: The line is the exact curve; triangles and squares mark the
 curves calculated, respectively, by
the approximations\Iii and \Viii .

\bye